\documentclass[12pt]{iopart}

\usepackage{ntheorem}

\usepackage{latexsym}
\usepackage{amsfonts}
\usepackage{amssymb}

\usepackage{citesort}
\usepackage{psfrag}

\usepackage{color}

\usepackage{url}
\usepackage{cite}

\usepackage[mathscr]{eucal}

\usepackage{graphicx}
\usepackage{epsfig}

\usepackage{epstopdf}

\newcommand{\mCpp}{{\mcC}^+_p}
\newcommand{\dgamma}{\dot \gamma}

\newcommand{\zalpha}{\mathring \alpha}

\newcommand{\ztz}{\theta_0}
\newcommand{\mybar}{}

\newcommand{\zGamma}{\mathring{\Gamma}}

\newcommand{\mcL}{{\mycal L}}

\newcommand{\chr}[3]{\Gamma^{#1}_{#2#3}}
\newcommand{\pu}{\partial_u}
\newcommand{\pr}{\partial_r}
\newcommand{\p}[1]{\partial_{#1}}
\newcommand{\guu}{g_{uu}}
\newcommand{\gur}{g_{ur}}
\newcommand{\grr}{g_{rr}}

\newtheorem{theorem}{Theorem}[section]

\newtheorem{Proposition}[theorem]{\sc Proposition}

\newtheorem{Remark}[theorem]{\sc Remark}

\newtheorem{Theorem} [theorem] {\sc  Theorem\rm}

\newtheorem{Lemma} [theorem] {\sc  Lemma\rm}

\theoremsymbol{\ensuremath{\diamondsuit}}
\newcommand{\fcoco}{\small}
\theorembodyfont{\fcoco}\theoremseparator{}\theoremindent0.5cm

\theoremstyle{nonumberplain}\theorembodyfont{\fcoco}
\theoremseparator{}\theoremindent0.5cm

{\catcode `\@=11 \global\let\AddToReset=\@addtoreset}
\AddToReset{equation}{section}
\renewcommand{\theequation}{\thesection.\arabic{equation}}

{\catcode `\@=11 \global\let\AddToReset=\@addtoreset}
\AddToReset{figure}{section}
\renewcommand{\thefigure}{\thesection.\arabic{figure}}

{\catcode `\@=11 \global\let\AddToReset=\@addtoreset}
\AddToReset{table}{section}
\renewcommand{\thetable}{\thesection.\arabic{table}}

\newcommand{\ubt}[2]{\underbrace{#1}_{#2}}
\newcommand{\ub } {\underbrace }

\newenvironment{proof}[1][Proof]{\textsc{#1.} }{\ \rule{0.5em}{0.5em}}

\DeclareFontFamily{OT1}{rsfs}{} \DeclareFontShape{OT1}{rsfs}{m}{n}{
<-7> rsfs5 <7-10> rsfs7 <10-> rsfs10}{}
\DeclareMathAlphabet{\mycal}{OT1}{rsfs}{m}{n}

\newcommand{\msqzh}{\sqrt{\det \zh_{EF}}}%
\newcommand{\zlambda}{\mathring \lambda}%
\newcommand{\zgamma}{\mathring \gamma}%
\newcommand{\zh}{\mathring h}%
\newcommand{\comp}[1]{}
\newcommand{\mMLI}[1]{}
\newcommand{\MLI}[1]{}

\newcommand{\beaa}{\begin{eqnarray*}}
\newcommand{\eeaa}{\end{eqnarray*}}

\newcommand{\bel}[1]{\begin{equation}\label{#1}}
\newcommand{\bea}{\begin{eqnarray}}
\newcommand{\bean}{\begin{eqnarray}\nonumber}
\newcommand{\beal}[1]{\begin{eqnarray}\label{#1}}
\newcommand{\eeal}[1]{\label{#1}\end{eqnarray}}
\newcommand{\eea}{\end{eqnarray}}

\newcommand{\Eq}[1]{Equation~\eq{#1}}

\renewcommand{\proof}{\noindent {\sc Proof:\ }}
\newcommand{\be}{\begin{equation}}
\newcommand{\eeq}{\end{equation}}
\newcommand{\ee}{\end{equation}}
\newcommand{\beqa}{\begin{eqnarray}}
\newcommand{\eeqa}{\end{eqnarray}}
\newcommand{\beqan}{\begin{eqnarray*}}
\newcommand{\eeqan}{\end{eqnarray*}}
\newcommand{\ba}{\begin{array}}
\newcommand{\ea}{\end{array}}

\newcommand{\mcM}{{\mycal M}}
\newcommand{\mcD}{{\mycal D}}
\newcommand{\hmcM}{\,\,\,\widehat{\!\!\!\mcM}}

\newcommand{\R}{\mathbb R}

\newcommand{\mcA}{\mycal A}
\newcommand{\mcC}{\mycal C}
\newcommand{\mcN}{\mycal N}

\newcommand{\mysign}{\varepsilon}

\newcommand{\nullhyp}{\mcN}

\newcounter{mnotecount}[section]

\newcommand{\mnote}[1]
{\protect{\stepcounter{mnotecount}}$^{\mbox{\footnotesize
$
\bullet$\protect\themnotecount}}$ \marginpar{
 \scriptsize
\raggedright \em $\!\!\!\!\!\!\,\bullet$\protect\themnotecount:
#1} }

\renewcommand{\themnotecount}{\thesection.\arabic{mnotecount}}

\newcommand{\ptc}[1]{\mnote{{\bf ptc:}#1}}
\renewcommand{\ptc}[1]{\mnote{ #1}}
\renewcommand{\ptc}[1]{}

\newcommand{\ptcx}[1]{}

\newcommand{\eq}[1]{(\ref{#1})}

\newcommand{\qed}{\hfill$\Box$ \medskip}

\begin{document}

\title{The light-cone theorem}
\author{
Yvonne Choquet-Bruhat$^{1}$,
Piotr T. Chru\'{s}ciel$^{2,3}$ and Jos\'{e} M.
Mart\'{\i}n-Garc\'{\i}a$^{4,5}$}
\address{$^1$ Acad\'emie des Sciences, Paris}
\address{$^2$ F\'{e}d\'{e}ration Denis Poisson, LMPT, Tours}
\address{$^3$ Hertford College and Oxford Centre for
Nonlinear PDE, University of Oxford}
\address{$^4$ Laboratoire Univers et Th\'eories, CNRS, Meudon, and
Universit\'e Paris Diderot}
\address{$^5$ Institut d'Astrophysique de Paris, CNRS, and
Universit\'e Pierre et Marie Curie}

\begin{abstract}
We prove that the area of cross-sections of light-cones, in
space-times satisfying suitable energy conditions, is smaller
than or equal to that of the corresponding cross-sections in
Minkowski, or de Sitter, or anti-de Sitter space-time. The
equality holds if and only if the metric coincides with the
corresponding model in the domain of dependence of the
light-cone.
\end{abstract}


\maketitle

\section{Introduction}

It is a well known fact in general relativity that gravitation
tends to focus null geodesics; this fact lies at the heart of,
e.g., the singularity theorems of Hawking and
Penrose~\cite{HawkingPenrose}. In this work we wish to point a
simple and striking illustration of this fact, which seems to
have been overlooked in the literature, concerning the area of
cross-sections of light-cones: We prove that such
cross-sections, in  space-times satisfying the Einstein
equations with vanishing cosmological constant $\Lambda$, and
with the energy-momentum satisfying the dominant energy
condition, are smaller than the corresponding areas of
cross-sections of light-cones in Minkowski space-time.
Moreover, under supplementary restrictions on the
energy-momentum tensor, equality of areas for a cross-section
$S$ implies that the space-time is Minkowskian in the domain of
dependence of that part of the light-cone which lies between
the vertex and the cross-section $S$. A similar result holds
when $\Lambda \ne 0$: in the statement just given one needs to
replace the Minkowski space-time by the de Sitter or anti-de
Sitter space-time. The precise statements can be found in
Section~\ref{STt}.

The idea of the argument is to show, using the dominant energy
condition, that the expansion of the light-cone is smaller than
that of the model space; this implies the area inequality. The
rigidity part of our statement is based on an analysis, closely
following that in~\cite{RendallCIVP}, of the associated
characteristic Cauchy problem; see
also~\cite{DossaAHP,F1,YvonneCIVP,RendallCIVP2,DamourSchmidt,PenroseCIVP}
and references therein.

\section{The theorem}
 \label{STt}

Consider an $(n+1)$--dimensional space-time $(\mcM,g)$, $n\ge
2$, satisfying the \emph{dominant energy-condition},
\bel{rdec}
 T_{\mu\nu}X^\mu Y^\nu \ge 0 \ \mbox{for all future oriented
 timelike vectors $X$ and $Y$.}
\ee
This will be the only condition needed for our comparison
result. However, to obtain rigidity, more conditions will be
needed. We shall say that the \emph{rigid dominant energy
condition holds at $q\in \mcM$} if \eq{rdec} holds, together
with the implication:
\bea
 &T_{\mu\nu}X^\mu X^\nu = 0 \ \mbox{for some
causal vector  $X$ at $q $}\ \Longrightarrow \
 T_{\mu\nu}X^\nu=0 \ \mbox{at $q$}. \phantom{xxx}
\eeal{rdec2}
(It is well known that the implication is always true for
timelike vectors by \eq{rdec} (compare Appendix~\ref{ALDEC}),
so this is only a restriction for null $X$'s.) We note a
related condition used by Galloway and Solis~\cite{GS}  (see
condition (C) in Section~4 of that last reference), also in a
null rigidity context.

General relativistic fluids with timelike flow vector $u^\mu$,
with $0\le |p|\le \rho$, and with an equation of state which
excludes the possibility $p=-\rho$ except when $\rho= 0$,
provide energy-momentum tensors satisfying \eq{rdec2}
everywhere.
Another example is provided by the energy-momentum $T_{\mu\nu}=
\rho \ell_\mu \ell_\nu$, where $\rho \ge 0$ and $\ell_\mu$ is
null.

Examples of energy-momentum tensor satisfying the dominant
energy condition and which do  \emph{not} satisfy \eq{rdec2}
are given by $T_{\mu\nu} = -\rho g_{\mu\nu}$, $\rho\ge 0$,%
\footnote{Our signature is $(-,+\ldots,+)$.}
or by  massless scalar fields, or by the Maxwell
energy-momentum tensor, as discussed in Appendix~\ref{Ardc}.

There is, however, a version of \eq{rdec2} which applies to
both massless scalar fields and Maxwell fields; see
Propositions~\ref{PEM} and \ref{PScalar} below; we emphasize
that the argument there is \emph{non-local} (as it requires
integration), and \emph{non-algebraic} (as it makes use of the
field equations): To define this, let $\ell$ be a field of null
tangents to a null hypersurface $\mcN$. We shall say that the
\emph{rigid dominant energy condition holds on $\mcN$} if
\eq{rdec} holds together with the implication:
\bea
 &T_{\mu\nu}\ell^\mu \ell^\nu = 0 \ \mbox{on $ \mcN$}\ \Longrightarrow \
 T_{\mu\nu}\ell^\nu=0 \ \mbox{on $ \mcN$}. \phantom{xxx}
\eeal{rdec3}

\begin{figure}[t]
\begin{center} {
\psfrag{gamltwo}{\Huge $\gamma_{\ell_2} $}
\psfrag{gamlone}{\Huge $\gamma_{\ell_1} $}
\psfrag{Tv}{\Huge $T$ }
\psfrag{point}{\Huge $p$ }
\psfrag{Cdes}{\Huge $\mcC(s)$ }
\psfrag{Ades}{\Huge $\mcA(s)$ }
\resizebox{3in}{!}{\includegraphics{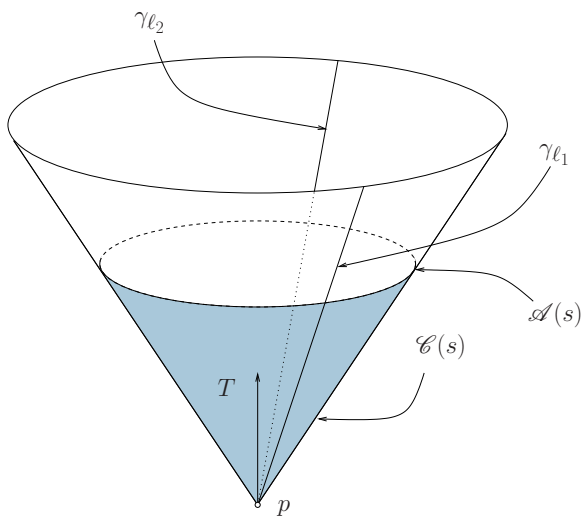}} }
\caption{\label{Fcones}
The cross-section $\mcA(s)$ of the light-cone $\mCpp$;  $\mcC(s)$ is the shaded blue region.
Two generators $\gamma_{\ell_1}$
and $\gamma_{\ell_2}$ are also shown.}
\end{center}
\end{figure}

Let $p\in \mcM$ and let $\mCpp$ be the future light-cone
emanating from $p$. Let $T$ be any unit timelike vector at $p$,
and normalize all null vectors $\ell$ at $p$ by requiring that
$g(\ell,T)=-1$. This defines an affine parameter, denoted by
$s$, on the future null geodesics $s\mapsto\gamma_\ell(s)$ with
$\gamma_\ell(0)=p$ and with initial tangent $\ell$. Let
$\mcA(s)$ denote the $(n-1)$--dimensional surface reached by
these geodesics after affine time
$s$:
\bel{20II.5}
 \mcA(s)=\{\gamma_\ell(s)\}\subset \mCpp
 \;,
\ee
where the vectors $\ell$ run over all null future vectors at
$p$ normalized as above; see Figure~\ref{Fcones}.
We denote by $\mcC(t)$  the subset of the light-cone covered by
all the geodesics up to affine time $t$:
\bel{20II.6}
 \mcC(t) = \cup_{0\le s\le t} \mcA(s)
 \;.
\ee
Note that $\gamma_\ell(s)$ might not be defined for all $s$.
Further, $\mcA(s)$ might not be a smooth surface. However, for
every point $p$ there exists a maximal $s_0>0$ such that
$\mcA(s)$ is defined and smooth for all $s<s_0$. We restrict
ourselves to $s<s_0$, though it is rather clear that this can
be relaxed using the methods of~\cite{ChDGH}; we have, however,
not attempted to verify all details of that.

Let $|\mcA(s)|_g$ denote the area of $\mcA(s)$. So for the
Minkowski metric, which we denote by $\eta$,  we have
$$
 |\mcA(s)|_\eta = \omega_{n-1} s^{n-1}
 \;,
$$
where $\omega_{n-1}$ is the area of the unit round sphere in
$\R^n$.

\bigskip

We consider metrics satisfying the  Einstein equations with
cosmological constant $\Lambda \in \R$ and sources. We assume
smoothness of the metric for simplicity, though our result
 can be proved under weaker differentiability conditions:

\begin{Theorem}
 \label{TCone}
Let $({\mycal M},g)$ be a smooth globally hyperbolic
space-time, {solution of the Einstein equations with
energy-momentum tensor} satisfying the dominant energy
condition.
We restrict attention to $s$ such that $\mcC(s)$ lies within
the domain of injectivity of the exponential map at $p$. Then:

\begin{enumerate}
 \item The area $|\mcA(s)|_g$ satisfies the inequality
\bel{4XII8.5} |\mcA(s)|_g\le |\mcA(s)|_\eta \;.
\ee
\item Let equality be attained at some  $s=s_2$. If
    \emph{either}
\begin{enumerate}
        \item the rigid dominant energy holds at
            $\mcC(s_2)$, \emph{or}
\item the energy-momentum tensor is traceless,
\end{enumerate}
then the domain of dependence of $\mcC(s_2)$ is isometric
to the corresponding domain of dependence in Minkowski  or
(anti) de Sitter space-time.
\end{enumerate}
\end{Theorem}

\proof Let $\theta$ denote the rate of change of area along the
null geodesic generators of $\mCpp$, and let $\sigma$ denote
the shear of $\mCpp$ (see,
e.g.,~\cite{galloway-nullsplitting}). 
Let $\gamma$ be such a generator, and recall the Raychaudhuri
equation in
space-time dimension $n+1$~\cite{galloway-nullsplitting} (note
that the rotation term vanishes because our family of null
geodesics forms a hypersurface)
\bel{Cthm1}
   \frac{d\theta}{ds}
 = - \sigma^{AB}\sigma_{AB} - \frac 1 {n-1} \theta^2  -R_{  \sigma  \mu}\dgamma^\sigma \dgamma^\mu
    \;.
\ee
Here $s$ is an affine parameter along the generators: $\nabla_{ \dot \gamma} \dot \gamma = 0$.%
\footnote{For the proof of rigidity we will be using a
coordinate system $(u,r,x^A)$, where $s=r$, with a  wave-map
condition imposed on the extension of the coordinates away from
the light-cone. However, no such condition is needed for the
comparison argument.}

Before giving a detailed proof, it might be useful to present
an outline: Let $\theta_0$ denote the expansion of a light-cone
in Minkowski space-time:
$$
 \ztz:= \frac{n-1}s
 \;;
$$
then $\ztz$  satisfies \eq{Cthm1} with vanishing Ricci tensor
and $\sigma$. Since $\theta$  approaches $(n-1)/s$ as the tip
of the light-cone is approached, a comparison argument using
\eq{Cthm1} shows that $\theta$ is smaller than its Minkowskian
value. This, subsequently, implies the area inequality.
Equality holds on $\mcA(s_2)$ if and only if $\sigma$ and $R_{
\sigma \mu}\dgamma^\sigma \dgamma^\mu$ vanish along all
geodesic generators of $\mCpp$ until these generators reach
$\mcA(s_2)$, i.e., on $\mcC(s_2)$. When the rigid dominant
energy condition holds (in either its local or nonlocal form),
the usual energy calculation implies that the metric is vacuum
in the domain of dependence of $\mcC(s_2)$. Under the traceless
condition  a more detailed analysis is necessary. This,
together with the vanishing of $\sigma$ on $\mcC(s )$, is used
to show that the metric tensor takes the model-metric values on
$\mcC(s )$, and the result follows by uniqueness of solutions
of the characteristic initial value problem.

Let us pass now to the details of the above. Since $\ztz$
satisfies the equation
$$
 \frac {d\ztz} {ds} = - \frac {\ztz^2}{n-1}
 \;,
$$
from \eq{Cthm1} we have
\bean
 \frac {d (\theta-\ztz)}{ds} & = & \frac{
 \ztz^2 -\theta^2  } { n-1  }- {  \sigma_{AB}\sigma^{AB} -  R_{\mu\nu}\dot\gamma^\mu
 \dot \gamma^\nu}
\\
 & =  &   -\frac{(\theta-\ztz)^2 } {n-1}  -\frac{2 } {s}(\theta-\ztz)-{  \sigma_{AB}\sigma^{AB} -  R_{\mu\nu}\dot\gamma^\mu
 \dot \gamma^\nu}
\nonumber
\\
 &\le&   -\frac{2 } {s}(\theta-\ztz)-{  \sigma_{AB}\sigma^{AB} -  R_{\mu\nu}\dot\gamma^\mu
 \dot \gamma^\nu}
 \;.
 \label{3XII8.6}
\eea
Hence, for $s>s_1>0$,
\bel{4XII8.1}
  s^2(\theta-\ztz)(s )\le   s_1^2(\theta-\ztz)(s_1 )- \int_{s_1}^{s} (\sigma_{AB}\sigma^{AB}
  +  R_{\mu\nu}\dot\gamma^\mu
 \dot \gamma^\nu)s^2
  ds
 \;.
\ee
Now, for a smooth metric we have
\bel{4XII8.0}
 \theta= \frac{(n-1)+o(1)}s
\ee
for small $s$, so  we can pass to the limit $s_1\to 0$ to
obtain
\bel{4XII8.3}
  (\theta-\ztz)(s )\le    - \frac 1 {s ^2}\int_{0}^{s} (\sigma_{AB}\sigma^{AB}
  +  R_{\mu\nu}\dot\gamma^\mu
 \dot \gamma^\nu)s^2
  ds
 \;.
\ee
Since the dominant energy condition has been assumed to hold,
the right-hand-side of \eq{4XII8.3} is non-positive and we
conclude that
\bel{3XII8.15}
 \theta(s)\le \frac {n-1}{s}
\ee
as long as the geodesic exists. Furthermore, equality holds for
some $s_2>0$ if and only if
\bel{3XII8.11}
 \forall s   \ \,    \mbox{satisfying} \ 0<s<s_2\;,  \quad \sigma_{AB}=0=R_{\mu\nu}\dot\gamma^\mu
 \dot \gamma^\nu
 \;.
\ee

The area inequality follows from \eq{3XII8.15} in a standard
way, we give the details for completeness. In a coordinate
system adapted to the light-cone we can write the metric on the
cone in the form
\bel{Cthm3}
 g = -\alpha du^2 + 2 \nu_A dx^A du - 2 e^{2\beta} du dr +
  g_{AB} dx^A dx^B
 \;,
\ee
so that $\mCpp=\{q\in \mcM: u(q)=0\}$, where $r$ is an affine
parameter along the generators of $\mCpp$, vanishing at the
vertex, denoted by $s$ in the previous equations. A calculation
shows
\bel{Cthm4}
 \theta=   \frac {1} { \sqrt{ \det g_{AB} }}
 \partial_r(  \sqrt{ \det g_{AB} } )
 \;.
\ee
Let us denote by $\mathring g_{AB}dx^A dx^B$  the
$(n-1)$--dimensional corresponding metric arising on a
light-cone in the $(n+1)$--dimensional Minkowski space-time.
Then our analysis so far shows that
\[
\theta \equiv\partial
_{r}\log\sqrt{\det g_{AB} }\leq\theta_{0}\equiv\partial_{r}\log\sqrt
{\det\mathring g_{AB}}
 \;.
\]
Thus $\log(  {\det g_{AB} / {\det \mathring g_{AB}}})$ is
decreasing. By elementary considerations the quotient ${\det
g_{AB} / {\det \mathring g_{AB}}}$ tends to one as $r$ tends to
zero, and we conclude
\[
\log\sqrt{\det g_{AB} }\leq\log\sqrt{\det \mathring g_{AB}},{\rm \ \ hence
\ \ }\sqrt{\det g_{AB} }\leq\sqrt{\det \mathring g_{AB}}
 \;.
\]
The areas $|\mcA(r)|_g$ and $|\mcA(r)|_\eta$ are%
\[
|\mcA(r)|_g \equiv\int_{S^{n-1}}\sqrt{\det g_{AB} }\,dx^{2}%
\ldots dx^{n},\qquad |\mcA(r)|_\eta\equiv\int_{S^{n-1}}\sqrt{\det
\mathring g_{AB}}\,dx^{2}\ldots dx^{n}
 \;,%
\]
therefore
\[
|\mcA(r)|_g\le |\mcA(r)|_\eta
 \;,
\]
which establishes part 1. of the theorem.

\medskip

Assume, now, that equality in this last equation holds at
$s=s_2$. Equation \eq{3XII8.11} implies the vanishing of
$T_{\mu\nu}\dot \gamma^\mu \dot \gamma^\nu$ on $\mcC(s_2)$.

If we assume that the energy-momentum tensor $T$ satisfies the
rigid dominant energy condition, as in \eq{rdec2}, or the rigid
dominant energy condition on $\mcC(s_2)$, as in \eq{rdec3}, we
can conclude that $T_{\mu\nu}\dot \gamma^\nu$ vanishes on
$\mcC(s_2)$. The proof that the metric is vacuum in the domain
of dependence of $\mcC(s_2)$ is then standard, and proceeds as
follows:

Consider the manifold
$$
 \hmcM:= \mcM\setminus J^+(\mcA(s_2))
 \;,
$$
with the metric obtained from $g$ by restriction, still denoted
by $g$. Then $(\hmcM,g)$ is globally hyperbolic, with
\bel{7II.1}
 \mcD^+(\mcC(s_2),\hmcM) =
 \mcD^+(\mcC(s_2),\mcM)
 \;,
\ee
where $\mcD^+(\Omega,\mcM)$ denotes the domain of dependence of
an achronal set $\Omega$ within a space-time $(\mcM,g)$. The
equality in \eq{7II.1} means that the manifolds, equipped with
the obvious metrics, are isometric.

Let $t$ be any Cauchy time function on $\hmcM$, i.e., a time
function ranging over $\R$, the level sets of which are Cauchy
surfaces. Replacing $t$ by $t-t(p)$, we can without loss of
generality assume that $t(p)=0$. Let
\bel{5II.10x}
 E(s) = -\int_{\mcD^+(\mcC(s_2))\cap\{t=s\}} T^\mu{}_\nu n^\nu dS_\mu
 \;,
\ee
where $n^\mu$ is the field of future directed unit normals to
the level sets of $t$; $E$ is positive in our signature
$(-,+,\ldots,+)$. The divergence identity on the set bounded by
$ \mcC(s_2)\cap \{t\le s\}$ and $\mcD^+(\mcC(s_2))\cap\{t=s\}$
(compare \eq{5II.12} and Lemma~\ref{LDEC}) shows that, for any
time interval $[0,T]$, there exists a constant $C=C(T)$ such
that
\bel{13II1}
 E(s) \le C \int _0 ^s E(t) dt - \underbrace{\int_{\mcC(s)} T^\mu{}_\nu n^\nu dS_\mu}_{=0}
 \;,
\ee
where the boundary integrand vanishes by the rigid dominant
energy condition, being proportional to $ T^\mu{}_\nu n^\nu
\dot \gamma^\mu$. Since $E(s) $ approaches zero as $s$ tends to
zero, from Gronwall's lemma we obtain
$$
 \mbox{$E(s)=0$ for $0<s<s_2$.}
$$
Positivity of the integrand implies
\bel{4XII8.8}
 T_{\mu \nu} n^\mu n^\nu =0 \ \mbox{ on
 $\mcD^+(\mcC(s_2))$}
 \;.
\ee 

{}From \eq{4XII8.8} and Lemma~\ref{LDEC}, we conclude that an
energy-momentum tensor satisfying the rigid dominant energy
condition must vanish on every level set of $t$ within the
domain of dependence of $\mcC(s_2)$. As $\mcD^+(\mcC(s_2))$ is
covered by these level sets,  the vanishing of $T_{\mu\nu}$ on
$\mcD^+(\mcC(s_2))$ follows.

The proof of \eq{4XII8.8} for tensors that \emph{do not}
satisfy the rigid dominant energy condition requires more care.
In view of \eq{3XII8.11}, at this stage of the analysis we can
only conclude that
\bel{RrAz}
 R_{rr}= 8 \pi T_{rr}=0=8 \pi T_{rA}=R_{Ar}
\ee
on $\mcC(s_2)$. Indeed,  to see the vanishing of $T_{Ar}$, set
$\ell=\partial_r$. Then, by the dominant energy condition, the
vector field $T^{\mu}{}_\nu \ell^\nu
\partial_\mu $ is causal, and has vanishing scalar product with
$\ell $, hence is proportional to $\ell $. So $T^{\mu}{}_r
\partial_\mu $ is proportional to $\partial_r$; subsequently
$$
T_{Ar}=g_{A\mu}\ub{T^{\mu}{}_r}_{\mbox{\scriptsize $0$ unless $\mu =r$}}=\ub{g_{Ar}}_{0}T^{r}{}_r=0
 \;,
$$
as desired.\footnote{Actually, we can further show that \emph{a
traceless} $T_{\mu\nu}$ must vanish at the vertex of the
light-cone: for this, by continuity and \eq{3XII8.11} we find
that $T_{\mu\nu}\ell^\mu \ell^\nu=0$ at $p$ for every null
vector $\ell\in T_p\mcM$. By~\cite[Lemma~2.8]{poorman2},
$T_{\mu\nu}$ is proportional to the metric at $p$, and
tracelessness implies the claim. But this fact does not seem to
be useful in the analysis that follows.}

Let $x^\mu$ denote normal coordinates centered at $p$, let
$R>0$ denote the largest number so that the exponential map at
$p$ is a diffeomorphism from a truncated solid cone $
\Omega(R)\subset T_p\mcM$, defined as
$$
\Omega(R):=\{ 0\le x^0 \le R\;,\  r:=\sqrt{\sum (x^i)^2} < x^0
 \}
 \;,
$$
to its image in $\mcM$. Note that this image is included in
$\mcD^+(\mCpp(R))$ when the level sets of $x^0$ are timelike
within $\Omega(R)$.

If $\Lambda=0$, we let the functions $y^\mu$ be solutions of
the following characteristic Cauchy problem:
\beal{9XII8.2} &
 \Box_g y^\mu =0\;,
 &
\\
 &
 y^\mu|_{\mCpp(R)} = x^\mu
 &
 \;.
\eeal{9XII8.2a}
For non-zero $\Lambda$, we impose again the boundary conditions
\eq{9XII8.2a}, but we require instead that the map
$x^\alpha\mapsto y^\mu (x^\alpha)$  satisfies the wave-map
equation, with the (anti)-de Sitter metric in the target,
\bel{10II.10}
 \mathring g= -\left(1
  -\frac{2\Lambda}{n(n-1)} \,  r^2\right)dt^2 + \frac{dr^2} {1
  -\frac{2\Lambda}{n(n-1)} r^2} + r^2 \zh_{AB}dx^A dx^B
  \;,
\ee
where $\zh_{AB}dx^A dx^B$ is the round unit metric on
$S^{n-1}$. Thus, in both cases, the functions $y^\mu$ satisfy
the set of equations (see,
e.g.,~\cite[page 162]{YvonneBook})
\bel{10II.11}
 g^{\alpha \beta}
 \left(
  \partial_\alpha \partial _\beta y^\mu -\Gamma^\sigma_{\alpha \beta}  \frac{\partial y^\mu}{\partial x^\sigma}
  +\zGamma^\mu_{\nu \rho}\frac{\partial y^\nu}{\partial x^\alpha}\frac{\partial y^\rho}{\partial x^\beta}
  \right)
  =0
 \;,
\ee
where the $\zGamma^\mu_{\nu \rho}$'s are the Christoffel
symbols of $\mathring g$, except that \eq{10II.11} is linear
when $\Lambda=0$, and thus the solutions exist globally on the
domain of dependence of the smooth part of the light-cone,
while for $\Lambda \ne 0$ the solutions might exist only for
some neighborhood of the tip of the light-cone.

By~\cite[Theorem~5.4.2]{Friedlander75}
(compare~\cite{CagnacDossa}) the functions $y^\mu$ are smooth
up-to-boundary on $\mcD^+(\mCpp(R))$. Decreasing $R$ if
necessary, the functions $y^\mu$ form a smooth coordinate
system on  $\mcD^+(\mCpp(R))$. Let $g_{\mu\nu}$ denote the
components of the metric in the coordinates $y^\mu$, then the
$g_{\mu\nu}$'s are smooth up-to-boundary on $\mcD^+(\mCpp(R))$.
If we pass to a coordinate system so that
$$
 u: =y^0-|\vec y|\;,
$$
and $r=|\vec y|$, and where the $x^A$'s are local coordinates
on $S^{n-1}$, then the cone is given by the equation $u=0$, and
the metric on $\mCpp(R)$ takes the form \eq{Cthm3}:
\bel{9XII8.3}
 g = -\alpha du^2 + 2 \nu_A dx^A du -2  e^{2\beta} du\, dr +
 h_{AB} dx^A dx^B
 \;.
\ee
We emphasize that we \emph{do not} assume that the metric takes
the form \eq{9XII8.3} \emph{away} from $\{u=0\}$, so care must
be taken when $\partial_u$--derivatives are taken.
%

By definition~\cite{galloway-nullsplitting}, $\sigma_{AB}$ is
the trace-free part of
$$
 g(\nabla_A \partial_r,\partial_B) = g_{BC}\Gamma^C_{Ar} = \frac 12 \partial_r g_{AB}
 \;,
$$
so from the vanishing of $\sigma_{AB}$, and from the
explicit formula for $\theta=\ztz$ we obtain
\begin{equation}
\partial_{r} {h}_{AB}=\frac{2}{r} {h}_{AB}
 \quad
 \Longleftrightarrow
 \quad
\partial_{r}(r^{-2} {h}_{AB})=0
\;.
\end{equation}
Since $r^{-2} {h}_{AB}$ tends to the unit round metric
$\zh_{AB}$ on $S^{n-1}$ as $r$ tends to zero, we conclude that
$$
 h_{AB} = r^2 \mathring h_{AB}
 \;.
$$

We continue by showing that $\beta=0$. For this note that, by
definition of normal coordinates, $r$ is an affine parameter
along the geodesics generators of $\mcC_p^+$. So
$\nabla_{\partial_r} \partial_r =0$, which is equivalent to $
0=\Gamma^\mu_{rr}$. But
\bel{20II.1}
  \Gamma^\mu_{rr} = \delta^\mu_r(
2\pr\beta + \frac{1}{2}e^{-2\beta}\pu\grr)\;,
 \ \mbox{ and we conclude that } \  e^{-2\beta}\pu\grr = -4 \pr\beta
 \;.
\ee
We set
\bel{15II.1}
 \lambda^\mu:=  -
 g^{\alpha\beta}\Gamma^\mu_{\alpha\beta}
 \;,
\ee
\bel{15II.1x}
 \zlambda^\mu:= -
 g^{\alpha\beta}\zGamma^\mu_{\alpha\beta}\;.
\ee
%
The wave map condition $\lambda_r=g_{r\mu}\mathring \lambda^
\mu$ can be shown to read
$$
 \ubt{\frac{1}{2} h^{AB} \pr h_{AB}}{\frac{n-1}r}
 + \ubt{e^{-2\beta}  \pu\grr}{-4\pr \beta} \equiv \lambda_r = g_{r\mu}\mathring
 \lambda^ \mu \equiv \frac {n-1} r e^{2\beta}
 \;.
$$
Writing $y= e^{2\beta}$, this is the same as
\[
\partial_{r} y =  \frac{n-1}{2r}y(1-y)
 \;.
\]
Integrating, we obtain either $y\equiv  1$, or
$$
 y = \frac{C(x^A) r^{(n-1)/2}} {1+C(x^A) r^{(n-1)/2}}
 \;,
$$
for some function $C(x^A)$. But, in normal coordinates, $\beta$
approaches zero as $r$ goes to zero, and we conclude that
$y\equiv  1$; equivalently, $\beta \equiv 0$.

In Appendix~\ref{AppCheck} we show that the vanishing of
$R_{rA}$ is equivalent to
\begin{eqnarray} \nonumber
0
      &=& \frac{(n-2)(n-3)}{2r^2}\, \nu_A
       +  \frac{3n-5}{2r} \, \partial_r \nu_A
       +  \partial_r \partial_r \nu_A
\\
      &=& \frac{1}{r^{n-1}} \partial_r \left[ r^{n-1}\left(
             \partial_r \nu_A + \frac{n-3}{2r} \nu_A \right)\right]
             \;.
 \label{31I.11}
\end{eqnarray}
Integrating \eq{31I.11} in $r$ once we obtain, for some smooth
functions $\hat \nu_A=\hat \nu_A(x^B)$,
$$
 \hat \nu_A r^{1-n}= \partial_r \nu_A
   +\frac{n-3}{2r} \nu_A   = r^{\frac {-n+3}2}\partial_r (r^{{\frac {n-3}2}}\nu _A)
   \;.
$$
Integrating again, we conclude that there exist smooth
functions $ \mathring \nu_A(x^B)$ such that, for $n>1$,
\bel{bigproblem}
 \nu_A(r,x^B) = r^{\frac {3-n}2} \mathring \nu_A (x^B) - \frac
 2 { (n-1)} r^{ 2-n} \hat \nu_A (x^B)
 \;.
\ee
But from the definition of our coordinate system it is
elementary to show that $\nu_A$ approaches zero as $r\to 0$,
which implies that $\nu_A\equiv 0$.

We are ready now to  establish \eq{4XII8.8} for traceless
energy-momentum tensors. For this let
\bel{5II.11x}
 \Omega(s_*):= J^+(p)\cap \{t< s_*\}
  \;,
\ee
where $t=x^0$ is a normal coordinate. We define $s_*\le R$ to
be the largest number smaller than or equal to $s_2$ such that
$\Omega(s_*)$ lies within the domain of definition of normal
coordinates. Moreover, we assume that $\partial_t $ and $\nabla
t$ are timelike on $\Omega(s_*)$, and that the functions
$y^\mu$, defined as solutions of \eq{10II.11}, form a
coordinate system on  $\Omega(s_*)$. The proof of the vanishing
of $T_{\mu\nu}$, to be found in Appendix~\ref{AsEn}, is again
an energy calculation, using instead the energy functional
defined as
\bel{5II.10}
 E(s)=  -\int_{\mcD^+(\mcC(s_2))\cap\{t=s\}} T^\mu{}_\nu X^\nu dS_\mu
 \;,
\ee
where the normal--coordinates components of
$X=X^\mu\partial_\mu$ are, very roughly, of the form
\bel{6II.17} X^\mu =x^\mu
 \;.
\ee
This choice of $X^\mu$ ensures the vanishing of the boundary
term that arises on $\mcC(s_*)$ in the divergence identity
\eq{5II.12}. However, this leads to a difficulty because
$X^\mu$ is null at $\mcC(s)$, which implies that the integrand
of \eq{5II.10} does not control uniformly the energy as the
boundary $\mcC(s_*)$ of $\Omega(s_*)$ is approached. Thus, the
standard energy argument requires a careful reinspection. The
price to pay is the need to impose tracelessness of
$T_{\mu\nu}$. Moreover the argument does not guarantee that the
metric is vacuum throughout $\mcD^+(\mcC(s_2))$, but only on
$\mcD^+(\mcC(s_*))$,%
\footnote{Strictly speaking, the argument presented in the
Appendix~\ref{AsEn} only proves that the metric is vacuum in
$\Omega(s_*)$. But $\{t=s_*\}\cap J^+(\mcC(s_*))$ is a Cauchy
surface for  $\mcD^+(\mcC(s_*))$, so a standard argument proves
then that the metric is vacuum in $\mcD^+(\mcC(s_*))$.}
and we will   return to this issue at the end of the proof.


We let $s_*$ be the number defined in the paragraph after
\eq{5II.11x} when $T_{\mu\nu}$ is traceless, and we set
$s_*=s_2$ if the rigid dominant energy condition holds on
$\mcC(s_2)$. Since the metric is now vacuum on
$\mcD^+(\mcC(s_*))$, we have
\bel{20II.2}
 g^{AB}R_{AB}=2\Lambda
\ee
there. We shall use \eq{20II.2} to prove that
$\alpha=1-2\Lambda{r^2}/{n(n-1)}$ on $\mcC(s_*)$.

Recall that, at this stage,  on $\mcC(s_*)$ the metric takes
the form
\bel{Axcharmet}
 g = -\alpha du^2 - 2    du dr + r^2
 \zh_{AB} dx^A dx^B
 \;.
\ee
In Appendix~\ref{AppCheck2} we show that
\begin{equation}
g^{AB}R_{AB} =
   4\Lambda\frac{n+1}{n-1} +
   2 \partial_r\partial_r \alpha + \frac{3(n-1)}{r} \partial_r\alpha
   + \frac{(n-1)(n-2)}{r^2}(\alpha-1) .
\end{equation}
This, together with \eq{20II.2}, provides  a Fuchsian ODE for
$\alpha-1$, with characteristic exponents $\lambda$ which solve
the equation
$$
 2\lambda(\lambda-1) +3(n-1)\lambda + (n-1)(n-2)=0
 \;,
$$
and thus the solutions are
$$
 \alpha = 1
  -\frac{2\Lambda}{n(n-1)} \  r^2
  + \alpha_+(x^A) r^{\lambda_+} + \alpha_-(x^A)r^{\lambda_-}
 \;,
$$
where $\alpha_\pm$ are smooth functions on $S^{n-1}$, and
$$
 \lambda_\pm \in \left\{ \frac {1-n}2, 2-n\right\}
 \;.
$$
Since both characteristic exponents are negative, the only
regular solution is $\alpha \equiv
1-\frac{2\Lambda}{n(n-1)}r^2$.

We have therefore shown that $g_{\mu\nu}$ takes the Minkowski,
or (anti)-de Sitter form on $\mcC(s_*)$. Note that the  energy
argument above can be used to prove uniqueness of solutions of
the reduced Einstein equations, with the components of the
metric in the wave-map gauge prescribed on the light-cone,  in
the usual way (compare~\cite{DossaAHP,RendallCIVP2,PenroseCIVP}
and references therein). It follows that $g_{\mu\nu}$ equals
the corresponding reference metric on the domain of dependence
of $\mcC(s_*)$.

So, we have that $x^\mu=y^\mu$ on $\Omega(s_*)$, with
$g_{\mu\nu}= \mathring g_{\mu \nu}$ there. If $s_*< s_2$, then
one can repeat the argument of Appendix~\ref{AsEn} to obtain
the above conclusions on $\Omega(\hat s_*)$, for some $\hat
s_*$ satisfying $s_* < \hat s_* \le s_2$. Using this
observation, an easy open-closed argument shows that $s_*
=s_2$, which had to be established.
\qed

\medskip

For further reference we note the following result, which
follows immediately from \eq{3XII8.6} and \eq{4XII8.3}:

\begin{Proposition}
The expansion
 $\theta(s)$
will become negative along a generator $\gamma$ of $\mCpp$ at
some value of $s$ strictly smaller than $s_2$ whenever
\bel{9XII8.1}
  \int_{0}^{s_2} (\sigma_{AB}\sigma^{AB}
  +  R_{\mu\nu}\dot\gamma^\mu
 \dot \gamma^\nu)s^2
  ds \ge (n-1)s_2
 \;.
\ee
\qed
\end{Proposition}

Once $\theta(s)$ has become negative, standard arguments imply
that $\theta$ will diverge in finite time, so that either
$\gamma$ will be incomplete, or will leave $\dot J^+(p)$ in
finite time.

\appendix
\section{The rigid dominant energy condition on the null cone: Maxwell and scalar fields}
\label{Ardc}

We start by verifying:

\begin{Proposition}
 Both the Maxwell energy-momentum tensor  and the massless scalar
 field energy-momentum tensors satisfy the dominant energy
 condition.
\end{Proposition}

 \proof
 It suffices to show that if $n^\mu$ is unit and timelike, then
 $P_\mu:=T_{\mu\nu}n^\mu $ is causal. Now, in an orthonormal frame $e_\mu$
 with
 $n^\mu\partial_\mu =e_0$ we have, for the massless scalar
 field,
$$
 T_{00} = \frac 12 (e_0(\phi))^2 + \frac 12 \sum_i (e_i(\phi))^2\;,
 \quad
 T_{0i} =  e_0(\phi) e_i(\phi)
 \;,
$$
and the causal character of $P_\mu=T_{0 \mu }$ follows from
$a|\vec b| \le \frac 12 (a^2 + |\vec b|^2)$.

For the Maxwell field, further rotating the frame so that
$F_{0i} \sim \delta _i ^1$, it holds that
\beaa
 T_{00} &=& \frac 12      \sum_{ j} F_{0j}^2+ \frac 14   \sum_{i,j  } F_{ij}^2
 = \frac 12    F_{01}^2+ \frac 12   \sum_{ j} F_{1j}^2+ \frac 14   \sum_{i,j\ne 1 } F_{ij}^2\;,
\\
 T_{0i}
  &= &
      F_{01}   F_{i1}
 \;,
\eeaa
and the result follows as for the scalar field.
 \qed

 \medskip

Now we show that the scalar and Maxwell fields do not obey the
rigid dominant energy condition in its local form (\ref{rdec2})
at a point $q$. For a scalar field $\phi$, define $k_\mu \equiv
\partial_\mu\phi|_q$. Then the energy-momentum tensor at $q\in \mcM$
can be expressed as
\[
T_{\mu\nu} = k_\mu k_\nu - \frac{1}{2} |k|_g^2 g_{\mu\nu}
\]
For spacelike $k_\mu$ the associated tensor $T^\mu{}_\nu$ has
null eigenvectors (which are orthogonal to $k$) with nonzero
eigenvalue $-\frac{1}{2} |k|_g^2$, which implies that
$T_{\mu\nu}$ does not obey the rigid dominant energy condition
(\ref{rdec2}).

The Maxwell stress energy tensor of an electromagnetic field
is, whatever the space-dimension $n\ge 2$,
\[
T_{\alpha\beta}= F^{\lambda}{}_{\alpha}F_{\lambda\beta}-\frac{ 1 }%
{4}g_{\alpha\beta}F^{\lambda\mu}F_{\lambda\mu}\;.
\]
At any point at which  $F_{\mu\nu}$  is of the form
$Y_{[\mu}Z_{\nu]}$, for some \emph{spacelike} vectors $Y$ and
$Z$, there exist null vectors $l^\mu$ for which
$F_{\mu\nu}l^\nu = 0$. Such vectors are eigenvectors of
$T^{\alpha}{}_{\beta}$ with nonzero eigenvalue. This implies
that the Maxwell field does not obey the rigid dominant energy
condition (\ref{rdec2}).

Next, let $\mcC(s_2)$ be the subset of the future null cone
$\mcC^+_p$   defined by \eq{20II.5}-\eq{20II.6}. We have:

\begin{Proposition}
 \label{PEM}  In
space-times satisfying the Einstein-Maxwell field equations,
the rigid dominant energy condition holds on  $\mcC(s_2)$.
\end{Proposition}

\proof
We use a coordinate system in which the metric takes the form
\eq{9XII8.3}.
The condition $\mybar{T}_{\mu\nu} \ell^\mu \ell^\nu =0$ in
\eq{rdec3} reads in those coordinates
\[
\mybar{T}_{ r r }=\mybar{F}^{\lambda}{}_{ r }\mybar{F}_{\lambda r }=0
 \;,
\]
with, by antisymmetry, $\mybar{F}_{ r r }=\mybar{F}^{ u }{}_{ r
}=0 $. Hence
\[
\mybar{T}_{ r r }=\mybar{g}^{AB}\mybar{F}_{A r }\mybar{F}_{B r }=0
 \;,
\]
which implies
\[
\mybar{F}_{A r }=0,{\rm \ \ hence\ also \ }\mybar{F}^{A u}=0\;.
\]
Keeping in mind $g_{rr}=g_{rA}=0$, we obtain a direct,
alternative justification of \eq{RrAz}:
\[
\mybar{T}_{ rA}=\mybar{F}^{\lambda}{}_{ r }\mybar{F}_{\lambda A}=0\;.
\]

The Maxwell equation
  $dF=0$ shows that
\[
\partial_{ r }\mybar{F}_{AB}=0
 \;.
\]
Because of the polar character of the coordinates $x^A$,
regularity of $F$ at the vertex gives the vanishing of
$\mybar{F}_{AB}=0$ there, and hence everywhere.

The Maxwell equation
$$
 \partial_\mu (\sqrt{|\det  g_{\alpha\beta}|} F^{\mu u}) = 0
$$
reduces in our coordinates to
$$
 \partial_r (e^{-2\beta}\sqrt{\det  h_{AB} } F_{r u}) = 0
  \;.
$$
Since $e^{-2\beta}\sqrt{\det  h_{AB} } F_{r u}$ tends to zero
as $r\to 0$, we conclude that   $\mybar{F}_{ u  r }\equiv 0$.
Now (recall that
$\mybar{g}_{ u  r }=-e^{2\beta}$ and $\mybar{F}_{A r }=\mybar{F}^{A u}=0)$,%
\[
\mybar{T}_{ u  r }=F^{ u }{}_{ u }F_{ u  r }+\frac{ e^{2\beta} }{2}F^{ u  r }F_{ u  r }+\frac{ e^{2\beta} }{4}\mybar{F}%
^{AB}\mybar{F}_{AB}
 \;,%
\]
and so
\[
\mybar{T}_{ u  r }=0\;.
\]
Hence $T_{\mu\nu}\ell^\mu$ vanishes on $\mcC(s_2)$, as desired.
\qed

\medskip

Similarly, we have

\begin{Proposition}
 \label{PScalar}  In
space-times  satisfying the Einstein --- massless scalar field
equations, the rigid dominant energy condition holds on
$\mcC(s_2)$.
\end{Proposition}

\proof In this case
\[
T_{\alpha\beta}=\partial_{\alpha}\phi\,\partial_{\beta}\phi-\frac{1}%
{2}g_{\alpha\beta}\partial_{\lambda}\phi\nabla^{\lambda}\phi
 \;.
\]
Hence
\[
\mybar{T}_{rr}=(\partial_{r}\mybar{\phi})^{2},
\]
and $\mybar{T}_{rr}=0$ implies $\partial_{r}\mybar{\phi}=0$. So
$\phi$ is constant on $\mcC(s_2)$; uniqueness of solutions of
the wave equation implies that $\phi$ is constant in the domain
of dependence of $\mcC(s_2)$, and so $T_{\mu\nu}$ vanishes
there.
\qed
 \section{The dominant energy condition and its consequences}
 \label{ALDEC}
\newcommand{\myN}{n}

Let us denote $\sqrt{|g(Z,Z)|}$ by  $|Z|_g$. Given a timelike
vector $\myN $, let us denote by  $|Z|_{g,\myN }$  the square
root of
\bel{31I.5}
 g(Z,Z) + 2 \frac{g(Z,\myN) ^2}{|g(\myN ,\myN) |} \ge 0
 \;.
 \ee
Note that $|\myN |_{g}=|\myN |_{g,\myN }$, and also
$|Z|_{g}=|Z|_{g,\myN }$ when $Z$ is orthogonal to $\myN $.

\medskip

We recall a well known result, which we prove for completeness:

\begin{Lemma}
 \label{LDEC}
 Suppose that a symmetric two-covariant tensor $T$ satisfies the
dominant energy condition \eq{rdec}, and let   $\myN $ be a
timelike vector.%
 \footnote{We hope that the clash of notation with the space-dimension $n$, as used elsewhere in this paper,
 will not lead to confusions.}
Then for any vectors $W,Z$ we have
\bel{LDEC1}
 |T(W,Z)| \le   \frac{ |W|^2_{g,\myN }+ |Z|^2_{g,\myN } }{ |\myN |_g^2 } T(\myN ,\myN)
 \;.
\ee
Furthermore, for any causal vector $X$ we also have
\bel{LDEC2}
  T(X,X) \le   \frac{ 2 |X| _{g,\myN } }{  |\myN |_g  }  T(X,\myN)
 \;.
\ee
\end{Lemma}

\begin{Remark}
 \label{RLDEC}
{\rm
Denoting by $|T |_{g,\myN }$ the norm of $T$ with respect to
the Riemannian metric associated with the quadratic form
\eq{31I.5}, \eq{LDEC1} implies
\bel{LDEC3x}
 |T |_{g,\myN } \le \frac 2 {|\myN |_g^2 } T(\myN ,\myN)
 \;.
\ee
}
\end{Remark}

\proof
 Let, first $W$ be
orthogonal to $\myN $. As $|W|_{g,\myN } =|W|_g$, the vectors
$W_\pm:= |W|_{g,\myN } \myN \pm |\myN |_g W $ are null
consistently time-oriented, thus
$$
 0\le  T(W_+,W_-)   = |W|^2_{g,\myN } T(\myN ,\myN)  -|\myN |^2_g T(W,W)
 \;,
$$
giving, for  $W \perp \myN $,
\bel{31I.1}
  T(W,W) \le \frac{|W|^2_{g,\myN } }{ |\myN |^2_g} T(\myN ,\myN)
   \;.
\ee
Adding the two equations obtained by writing explicitly
$T(W_+,W_+)\ge 0$ and $T(W_-,W_-)\ge 0$ gives
\bel{31I.7}
  T(W,W) \ge -\frac{|W|^2_{g,\myN } }{ |\myN |^2_g} T(\myN ,\myN)
  \quad
  \Longrightarrow\quad
  |T(W,W)| \le \frac{|W|^2_{g,\myN } }{ |\myN |^2_g} T(\myN ,\myN)
   \;.
\ee
We also have,
$$
 0\le T(\myN ,W_\pm) = T(\myN , |W|_g \myN)  \pm T(\myN ,|\myN |_g W)
 \;,
$$
giving, again for   $W \perp \myN $,
\bel{31I.2}
 | T(W,\myN)  |\le \frac{|W| _{g,\myN } |\myN | _{g,\myN } }{ |\myN |^2_g} T(\myN ,\myN)
   \;.
\ee
Next, if both $W$ and $Z$ are orthogonal to $\myN $, using
\eq{31I.7} we find
\bean
 |T(W,Z) |&= &\frac 14 \left|  T(W+Z,W+Z) -  T(W-Z,W-Z) \right|
\\
 & \le  &   \frac{|W +Z| _{g,\myN }^2 +|W-  Z|
_{g,\myN }^2 }{4 |\myN |^2_g} T(\myN ,\myN)
 \nonumber
\\
 & =  & \frac{|W | _{g,\myN }^2 +|  Z|
_{g,\myN }^2 }{2 |\myN |^2_g} T(\myN ,\myN)
 \;.
\eeal{31I.3}
Finally, for general vectors $W$ and $Z$ we can write
$$
 W = w\frac {\myN  }{ |\myN | _g}+ W^\perp\;, \quad
 Z = z\frac {\myN }{ |\myN | _g}+ Z^\perp\;,
$$
with both $W^\perp$ and $Z^\perp$ orthogonal to $\myN $. Then
$$
 |W | _{g,\myN }^2 = w^2 + |W ^\perp| _{g,\myN }^2\;,
 \quad
 |Z | _{g,\myN }^2 = z^2 + |Z ^\perp| _{g,\myN }^2\;,
$$
and, from what has been said so far,
\beaa | T(W,Z) | &= & \left| \frac{wz}{ |\myN |^2_g}T(\myN
,\myN) +
 \frac{w }{ |\myN | _g}T(\myN ,Z^\perp)
 + \frac{z }{ |\myN | _g}T(\myN ,W^\perp) +T(W^\perp,Z^\perp)
 \right|
\\
 & \le & \frac{|wz| + |w Z^\perp | _{g,\myN }
 + |z W^\perp | _{g,\myN }
 + \frac 12(|W ^\perp |_{g,\myN }^2
  +  |Z^\perp |^2 _{g,\myN } )}{ |\myN |^2_g}
  T(\myN ,\myN)
\\
 & \le & \frac{w^2+z^2
 + |W^\perp  |_{g,\myN }^2
  +  |Z^\perp |^2 _{g,\myN } }{ |\myN |^2_g}
  T(\myN ,\myN)
\\
 & = & \frac{  |W |_{g,\myN }^2
  +  |Z|^2 _{g,\myN }  }{  |\myN |^2_g}
  T(\myN ,\myN)
 \;.
\eeaa
This proves \eq{LDEC1}.

For \eq{LDEC2}, set $Z^\mu = - T^\mu{}_\nu  X^\nu$; the
dominant energy condition implies that $Z^\mu$ is causal future
directed. Let $e_a$, $a \in \{0,\ldots,n\}$, be any orthonormal
frame such that $\myN  = |\myN |_g e_0$, and let $X^a$ denote
the components of $X$ in this frame, thus $X=X^a e_a$,
similarly for $Z^a$. Then \eq{LDEC2} is equivalent to
\bel{LDEC3}
  -g(Z,X) \le  \frac{ 2 |X| _{g,\myN } }{ |\myN |_g  }(- g(Z,\myN) )
 \;.
\ee
Now, since both $Z$ and $X$ are causal and future directed we
have $|\sum_i X^i Z^i| \le Z^0 X^0$, so
\beaa
 -g(Z,X) &=& Z^0 X^0 -\sum_i X^i Z^i \le 2 Z^0 X^0 = 2
 \frac {X^0}{\myN ^0} (-g(Z,\myN) )
\\
 &=& 2  \frac {X^0}{|\myN |_g} (-g(Z,\myN) )
 \;,
\eeaa
and \eq{LDEC2} follows.
\qed

We note that the constant in the Lemma~\ref{LDEC} is optimal,
with the inequality becoming an equality when $Z$ is null, when
$T_{\mu\nu} = Z_{(\mu}\myN_{\mu)}$, and when $X^0=Z^0$,
$X^i=-Z^i$.

\section{$R_{rA}$}
 \label{AppCheck}
In this appendix we calculate the components $R_{rA}$ of the
Ricci tensor of a metric which on a null hypersurface
$\nullhyp=\{u=0\}$ takes the form
\bel{xCthm3x}
 g =  - \alpha du^2+ 2 \nu_A dx^A du +2\mysign du dr +
 \ub{ r^2 \zh_{AB}}_{ h_{AB}}dx^A dx^B
 \;.
\ee
Here we allow  $\mysign =\pm 1$, according to whether a future
($\mysign =-1$) or a past ($\mysign=1$) light-cone is
considered. We emphasize that the above form of the metric is
only assumed at $\{u=0\}$, so all the $g_{\mu\nu}$'s are
allowed a priori to be non-zero away from $\nullhyp$; similarly
for their derivatives.

The equations in this appendix, and in appendix
\ref{AppCheck2}, have been checked with the {\em xAct} system
for tensor computer algebra \cite{xAct}.

Writing $g^\sharp$ for the inverse metric, we  have
\bel{xAinvcharmetx}
 g^\sharp = \psi\partial^2_r + 2 \mu^A \partial_r \partial_A
 +2\mysign   \partial_u \partial_r +\frac 1 {r^2}\mathring h^{AB}\partial_A
 \partial_B
  \;,
\ee
\bel{xxAinvrelchar}
 g^{rA}\equiv \mu^A =  -\mysign \frac 1 {r^2} \zh^{AB} \nu_B
 \;,
 \qquad g^{rr}\equiv \psi= \alpha+ \frac 1 {r^2} \zh^{AB}\nu_A \nu_B
 \;.
\ee

We reserve the symbols $\nu_A$, $\mu^A$, $\alpha$, $\psi$ and
$h_{AB}$ for objects defined on $\{u=0\}$, so that e.g.
$\partial_u \nu_A$ does not make sense (but $\partial_u g_{u
A}$ does, and might a priori be non-zero).

The Levi-Civita connection of the metric $h_{AB}$ will be
denoted as $D_A$ and will have Christoffel symbols
$\gamma^C_{AB}$ with respect to the derivative $\partial_A$.

All the equations that follow are on $\mcN$.

We have the following Christoffel symbols (the remaining ones
can be obtained by symmetry):
\begin{eqnarray}
\chr{u}{u}{u}
&=&
\frac{\mysign}{2} ( \pr\alpha + 2 \pu\gur )
\; ,
 \label{uuu}
\\
\chr{u}{u}{r}
&=&
\frac{\mysign}{2} \pu\grr
\; , \\
\chr{u}{r}{r}
&=&
0
\; , \\
\chr{r}{u}{u}
&=&
\frac{1}{2}\mu^A\p{A}\alpha
+ \frac{1}{2} \psi ( \pr\alpha + 2 \pu\gur )
+ \mu^A\pu g_{uA}
+ \frac{\mysign}{2}\pu\guu
\; , \\
\chr{r}{u}{r}
&=&
- \frac{\mysign}{2} \pr\alpha
+ \frac{1}{2}\mu^A\pr\nu_A
+ \frac{1}{2}\mu^A\pu g_{rA}
+ \frac{1}{2} \psi \pu\grr
\; , \\
\chr{r}{r}{r}
&=&
- \frac{\mysign}{2}\pu\grr
\; , \\
\chr{u}{A}{u}
&=&
\frac{\mysign}{2} (\pu g_{rA} - \pr\nu_A)
\; ,  \\
\chr{u}{A}{r}
&=&
0
\; , \\
\chr{r}{A}{u}
&=&
- \frac{\mysign}{2}\p{A}\alpha
+ \frac{1}{2}\mu^B (D_A\nu_B - D_B\nu_A + \pu g_{AB})
 \nonumber
\\
 &&
+ \frac{1}{2}\psi (\pu g_{rA} - \pr\nu_A)
\; , \\
\chr{r}{A}{r}
&=&
- \frac{\mysign}{2} ( \pu g_{rA} - \pr\nu_A )
+ \frac{1}{2} \mu^B \pr h_{AB}
\; , \\
\chr{u}{A}{B}
&=&
- \frac{\mysign}{2} \pr h_{AB}
\; , \\
\chr{r}{A}{B}
&=&
\frac{\mysign}{2} (D_A\nu_B+D_B\nu_A-\pu g_{AB})
- \frac{1}{2}\psi \pr h_{AB}
\; , \\
\chr{C}{u}{u}
&=&
\frac{1}{2} h^{CA}\p{A}\alpha
+ \frac{1}{2} \mu^C \pr\alpha
+ h^{CA} \pu g_{uA}
+ \mu^C \pu\gur
\; , \\
\chr{C}{u}{r}
&=&
  \frac{1}{2} h^{CA} (\pu g_{rA} + \pr\nu_A)
+ \frac{1}{2} \mu^C \pu\grr
\; , \\
\chr{C}{r}{r}
&=&
0
\; , \\
\chr{C}{A}{u}
&=&
  \frac{1}{2} \mu^C (\pu g_{rA} - \pr\nu_A )
+ \frac{1}{2} h^{BC}(D_A\nu_B - D_B\nu_A + \pu g_{AB} )
\; , \\
\chr{C}{A}{r}
&=&
\frac{1}{2} h^{BC} \pr h_{AB}
\; , \\
\chr{C}{A}{B}
&=&
\gamma^C_{AB}
- \frac{1}{2} \mu^C \pr h_{AB}
\; .
\end{eqnarray}

The traces of the Christoffel symbols read:
\begin{eqnarray}
\chr{\mu}{u}{\mu}
&=&
\mysign \pu\gur
+\frac{1}{2}\psi \pu\grr
+\mu^A \pu g_{rA}
+\frac{1}{2} h^{AB}\pu g_{AB}
\; , \\
\chr{\mu}{r}{\mu}
&=&
\frac{1}{2} h^{AB}\pr h_{AB}
\; , \\
\chr{\mu}{A}{\mu}
&=&
\frac{1}{2}h^{BC}\p{A}h_{BC}
\; .
\end{eqnarray}

Let $\lambda^\mu$ be defined by \eq{15II.1}, we have
\begin{eqnarray}
\lambda^u
&=&
- \pu\grr
+\frac{\mysign}{2} h^{AB} \pr h_{AB}
\; , \label{12I.1} \\
\lambda^r
&=&
- \mysign h^{AB}D_ B \nu_A
+ \pr\alpha
-\mu^A\mu^B \pr h_{AB}
+ \frac{1}{2} h^{AB} \psi \pr h_{AB}
\nonumber \\ &&
- 2 \mysign \mu^A \pr \nu_A
+ \frac{\mysign}{2} h^{AB} \pu g_{AB}
- \frac{\mysign}{2} \psi \pu\grr
\\
&=&
- \mysign h^{AB}D_ B \nu_A
+ \frac{\pr(\psi\sqrt{\det h_{EF}})}{\sqrt{\det h_{EF}}}
+ \frac{\mysign}{2} h^{AB} \pu g_{AB}
- \frac{\mysign}{2} \psi \pu\grr
\; ,\phantom{xxx} \\
\lambda^A
&=&
-h^{CD}\gamma^A_{CD}
+ \frac{1}{2} h^{BC} \mu^A \pr h_{BC}
- h^{AC} \mu^B \pr h_{BC}
\nonumber \\ &&
-\mysign ( h^{AB} \pr\nu_B + h^{AB} \pu g_{rB} + \mu^A \pu\grr )
\; , \\
\lambda_u
&=&
- h^{AB}\p{B}\nu_A
+ \mysign \pr\alpha
-\mu^A\pr\nu_A
\nonumber \\ &&
+\frac{1}{2} h^{AB}\pu g_{AB}
+\mu^A \pu g_{rA}
+\frac{1}{2} \psi \pu\grr
\; , \\
\lambda_r
&=&
\frac{1}{2} h^{AB} \pr h_{AB}
- \mysign \pu\grr
\; , \\
\lambda_A
&=&
- h^{BC}h_{AD}\gamma^D_{BC}
- \mu^B \pr h_{AB}
- \mysign ( \pr\nu_A + \pu g_{rA} )
\; .
\end{eqnarray}

We choose the metric \eq{10II.10} as model metric, expressed in
the following coordinate system:
\bean
 \mathring g &= &  -\underbrace{\left(1
  -\frac{2\Lambda}{n(n-1)} \,  r^2\right)}_{=:\zalpha}\underbrace{dt^2}_{(du - \frac \mysign \zalpha dr)^2} + \frac{dr^2} {1
  -\frac{2\Lambda}{n(n-1)} r^2} + r^2 \zh_{AB}dx^A dx^B
\\
 & = &   -\zalpha du^2 +2\mysign du dr +
  r^2 \mathring h_{AB} dx^A dx^B
  \;.
\eeal{10II.15}
Its non-vanishing Christoffel symbols are, up to symmetry,
\bean
 &
 \displaystyle
 \zGamma^u_{uu} = - \frac{2\mysign \Lambda r}{n(n-1)}
 \;, \quad
\zGamma^u_{BC} = -\mysign r \zh_{BC}
 \;,
 \quad  \zGamma^r_{uu} =  -\frac{2\Lambda r}{n(n-1)} \zalpha
 \;,
 &
\\
 &
 \displaystyle
\zGamma^r_{ur} = \frac{2\mysign \Lambda r}{n(n-1)}
 \;,
 \quad
 \zGamma^r_{BC} = -r \zh _{BC} \zalpha
 \;, \quad
 \zGamma^A_{Br}= \frac 1r \delta^{A}_{B}
 \;, \quad
\zGamma^A_{BC}= \zgamma^A_{BC}
 \;. \phantom{xxx}
 &
 \eeal{y13XI8.4}

We shall shortly assume that the metric $g$ satisfies the
\emph{wave-map conditions} (see,
e.g.,~\cite[Chapter~VI]{YvonneBook})
$$
\lambda^\mu = \zlambda^\mu
 \;,
$$
with $
 \zlambda^\mu$ defined in \eq{15II.1x}. We find
\bel{12I.2}
 \zlambda^u = -g^{\mu\nu}\zGamma^u_{\mu\nu} =  r\mysign g^{AB }\zh_{AB}=\mysign\frac {n-1}r
 \;,
\ee
\bel{12I.2x}
 \zlambda^r =- g^{\mu\nu}\zGamma^r_{\mu\nu} =\frac {n-1}r  -\frac{2(n+1) \Lambda
 r}{n(n-1)}
 \;,
\ee
\bea
 \zlambda^A &=& - g^{\mu\nu}\zGamma^A_{\mu\nu} = -2 g^{r B}\zGamma^A_{r B }
 -g^{BC}\zGamma^A_{BC} =
 - \frac 2 r  g^{r A}  - \frac 1 {r^2} \zh^{BC}\zgamma^A_{BC}
 \nonumber
\\
 &=&
  -\frac{2}{r} \mu^A
  + \frac 1 {r^2 \msqzh} \partial_B(\msqzh \zh^{AB})
 \;.
\eeal{31I.10}
Using $\lambda^u=\zlambda^u$, from \eq{12I.1} and \eq{12I.2} we
obtain
$$
 \partial_u  g_{rr} = 0
 \;, \ \mbox{hence also }\
 \partial_u  g^{uu} = 0
 \;.
$$
From $\lambda^A = \zlambda^A$ we deduce that
\beaa
 \partial_u g_{C r}& = & - \frac {(n-1)} r \nu_C - \partial_r \nu_C
 \;,
\eeaa
and finally $\lambda^r = \zlambda^r$ gives
\beaa \frac{1}{2} \zh^{AB}\partial_u g_{AB} &=&
\zh^{AB}D_A\nu_B -\mysign
\frac{\partial_r(r^{n-1}\psi)}{r^{n-3}} +(n-1)\mysign r
-\frac{2(n+1)\Lambda\mysign r^3}{n(n-1)} \;. \eeaa

Now,
\beaa
  R_{Ar}
 & =& \partial_\gamma
\Gamma^\gamma_{Ar}-\partial_r \Gamma^\gamma_{A\gamma} +
\Gamma^\gamma_{\sigma\gamma}\Gamma^\sigma_{Ar}-
 \Gamma^\gamma_{\sigma r}\Gamma^\sigma_{A\gamma}
 \;, \nonumber
 \eeaa
and from what has been said so far, in particular using the
harmonicity conditions, we obtain
\beaa
  R_{Ar}
  & =& \partial_u \Gamma^u_{Ar}
  +\partial_r {\Gamma^r_{Ar}}
  +\partial_B \Gamma^B_{Ar}
  -\partial_r \Gamma^\gamma_{A\gamma} +
\Gamma^\gamma_{u\gamma}\ub{\Gamma^u_{Ar}}_0+
 \underbrace{\Gamma^\gamma_{r\gamma}}_{ \Gamma^B_{rB} }\Gamma^r_{Ar}+
\Gamma^\gamma_{B\gamma}\Gamma^B_{Ar}
 \nonumber
\\
 &&
 - \ub{\Gamma^u_{u r}}_0\Gamma^u_{A u}-
 \ub{\Gamma^u_{r r}}_0\Gamma^r_{A u}-
 \ub{ \Gamma^u_{B r}}_0\Gamma^B_{A u}-
 \ub{\Gamma^i_{u r}\Gamma^u_{Ai}}_{\Gamma^B_{u r}\Gamma^u_{AB}}-
 \ub{\Gamma^i_{r r}}_0\Gamma^r_{Ai}-
 \Gamma^i_{B r}\Gamma^B_{Ai}
 \nonumber
\\
  & =& \partial_u \Gamma^u_{Ar}
  +\partial_B \Gamma^B_{Ar}
  -\partial_r \Gamma^u_{Au}
  -\partial_r \Gamma^B_{A B}  +
 \Gamma^B_{rB}\Gamma^r_{Ar}+
 \Gamma^\gamma_{B\gamma}\Gamma^B_{Ar}
 \nonumber
\\
 && - {\Gamma^B_{u
 r}\Gamma^u_{AB}}-
 \Gamma^r_{B r}\Gamma^B_{Ar}-
 \Gamma^C_{B r}\Gamma^B_{AC}
 \nonumber
 \;.
\nonumber
 \eeaa
We have:
\beaa
 \partial_u \Gamma^u_{Ar}
  & = & \frac 12 \partial_u\Big( g^{u \mu} ( \partial_A g_{\mu r} +
  \partial_r g_{\mu A} - \partial_\mu g_{Ar})\Big)
=
 \frac 12
 \partial_u g^{u  B}
   \partial_r g_{B A}
 \;.
\eeaa
On the null surface it holds that
\be
\partial_u g^{uB} = -\mysign h^{AB} \partial_u g_{rA}
                    -\mysign \mu^B \partial_u g_{rr}
                    \;,
\ee
so, using the harmonicity conditions, we are led to
\be
\partial_u g^{uB} = -\frac{n-1}{r} \mu^B + \mysign
h^{AB}\partial_r\nu_A\;. \ee
Hence
\be
\partial_u \Gamma^u_{Ar} =
\mysign \frac{1}{r^2} \left((n-1)\nu_A + r \partial_r \nu_A
\right) . \ee 
With some work, using the formulae derived so far, one
similarly obtains
\begin{equation}
\partial _{B}\mybar{\Gamma}_{rA}^{B}-\partial _{r}\mybar{\Gamma}_{AB}^{B}= -%
\frac{\mysign}{r}\partial _{r}\nu _{A}+\frac{\mysign}{ r ^{2}}\nu _{A}
 \;,
\end{equation}
\begin{equation}
-\partial _{r}\mybar{\Gamma}_{Au}^{u}= -\frac{\mysign }{2}\partial _{r}\{%
\mybar{\partial _{u}g_{rA}}-\partial _{r}\nu _{A}\}
 \;,
\end{equation}
\begin{equation}
\mybar{\Gamma}_{rA}^{r}\mybar{\Gamma}_{rB}^{B}
 =\mysign \frac{n-1}{r}\left\{\frac{1}{2}(\partial
_{r}\nu _{A}-\mybar{\partial _{u}g_{rA}})-\frac{1}{r}\nu _{A}\right\}
 \;,
\end{equation}
\begin{equation}
\mybar{\Gamma}_{rA}^{B}(\mybar{\Gamma}_{Bu}^{u}+\mybar{\Gamma}_{Br}^{r}+\mybar{\Gamma%
}_{BC}^{C})= \frac{1}{r}\mathring \gamma_{AC}^{C}
 \;,
\end{equation}
\begin{equation}
-\mybar{\Gamma}_{ru}^{B}\mybar{\Gamma}_{AB}^{u}= \frac{\mysign}{2r}(\partial
_{r}\nu _{A}+\mybar{\partial _{u}g_{rA}})
 \;,
\end{equation}
\begin{equation}
-\mybar{\Gamma}_{rB}^{r}\mybar{\Gamma}_{Ar}^{B}-\mybar{\Gamma}_{rC}^{B}\mybar{\Gamma}%
_{AB}^{C}= -\frac{\mysign}{2r}(\partial _{r}\nu _{A}-\mybar{\partial
_{u}g_{rA}})-\frac{1}{r}\mathring \gamma_{AB}^{B}
 \;.
\end{equation}

Adding, we are led to
\begin{eqnarray} \nonumber
\mysign R_{Ar}
      &=& \frac{(n-2)(n-3)}{2r^2}\, \nu_A
       +  \frac{3n-5}{2r} \, \partial_r \nu_A
       +  \partial_r \partial_r \nu_A
\\
      &=& \frac{1}{r^{n-1}} \partial_r \left[ r^{n-1}\left(
             \partial_r \nu_A + \frac{n-3}{2r} \nu_A \right)\right].
             \label{1II.1}
\end{eqnarray}\section{An energy inequality for traceless $T_{\mu\nu}$}
 \label{AsEn}
\newcommand{\zA}{z^A}%
\newcommand{\zB}{z^B}%
We let $x^\mu$, with $x^0\equiv t$, denote normal coordinates
centred on the vertex $p$ of the future light-cone; we restrict
consideration to the region where those are well defined.
Passing to a subset of the domain of normal coordinates if
necessary we can, and will, assume that $\partial_t$ and
$\nabla t$ are timelike. We will only consider metrics which
behave as in the proof of Theorem~\ref{TCone}: Thus, we assume
existence of a set of coordinates  $y^\mu$ which are required
to coincide with the normal coordinates $x^\mu$ on the
light-cone, and we assume that the map $x^\mu\mapsto y^\mu$ is
a smooth diffeomorphism in a neighbourhood of the future
light-cone of $p$. We let $u=y^0-|\vec y|$, $r=|\vec y|$, and
we denote by $\zA$ are angular coordinates parameterising the
unit vector $\vec y/r$. We denote by $(z^\mu)\equiv (u,r,\zA )$
these coordinates; by definition,  $\{u=0\}$ is $\mcC(s_*)$.
Furthermore we assume that, on $\mcC(s_*)$, the metric takes
the form \eq{Axcharmet},
\bel{16III.1}
 g = -\alpha du^2 - 2    du \, dr + r^2
 \zh_{AB} d\zA  d\zB
 \;.
\ee
Note that we write here $\zA$ for what is denoted by $x^A$
elsewhere in the paper since, to avoid confusions, in the
considerations below we reserve the symbol $x^\mu$ for normal
coordinates.

We will also need the hypothesis that $g_{uu}<0$ and that the
$u$-derivatives of the metric at the light-cone satisfy
$$
 \partial_u g_{rr}
  =0
 \;.
$$
As already pointed out, all those conditions will be the
satisfied by the wave-map coordinates from the main body of the
paper at the current stage of the argument. But we emphasize we
do not need to assume that the coordinates $y^\mu$ satisfy more
conditions than the ones just listed.

Consider a vector field $X$ which, near the light cone, equals
\bel{10IV.1}
 X=  u \partial_u +r\partial_r
 \;.
\ee
So, wherever $X$ takes this form,
$$
 g(X,X)= g_{uu} u^2 + 2g_{ur} u r + g_{rr} r^2
 \;.
$$
On the light-cone this vanishes, so that $X$ is null there.

Keeping in mind that $g_{uu}<0$ and $g_{ur}=-1$ at $\{u=0\}$,
for every $R>0$ there exists $u_0>0$ and $\epsilon>0$ such that
for $0\le r\le R$ and $0< u \le u_0$ we have $g_{ur} <0$ and
$g_{uu}<-\epsilon$. Since $\partial_u g_{rr}=0$ at $u=0$ we
further have, in the same ranges of $u$ and $r$, $|g_{rr}|\le C
u^2$. So the second term is negative, while there exists a
(possibly small) $r_0>0$ such that for $0< u \le u_0$ and $0\le
r\le r_0$ the first term dominates the third one, which shows
that $X$ is timelike in that region.

At $u=0$ we have
$$
 \partial_u \big(g(X,X)\big)|_{u=0}=   (-2  + r\partial_u g_{rr}) r
 \;.
$$
This shows that, reducing $u_0$ if necessary, $X$ is again
timelike in the range $r_0\le r \le R$ and $0<u\le u_0$, and
hence in the whole range $0\le r \le R$ and $0<u\le u_0$.

Since the set of future directed timelike vectors is convex,
for every $R>0$ one can interpolate in the region $u_0/2\le u
\le u_0$ between $X$ as given by \eq{10IV.1} and some vector
field timelike everywhere to obtain a vector field, still
denoted by $X$, which is timelike in the timelike future of $p$
and which takes the form \eq{10IV.1} for $0<r\le R$ and $0<u\le
u_0/2$.

\medskip

As $X$ is null at the light-cone, the integrand of \eq{5II.10}
does \emph{not} control all components of $T_{\mu\nu}$
\emph{uniformly} as one approaches the light-cone, and we need
to quantify that. So we start by showing that, for any
$T\in[0,\infty)$  there exists a constant $C>0$ such that for
$0<r<t\le T$ we have
%
\bel{Tnx}
 T_{\mu\nu} n^\mu X^\nu \ge   C^{-1}  (t-r)
 T_{0\nu}   n^\nu
 \;.
\ee
Note that for $u\ge u_0$,  the inequality follows immediately
from the fact that all three vectors $X^\mu$, $\partial_t$, and
$n^\mu$ are uniformly timelike there, and from \eq{LDEC2}. So
it remains to consider points for which   $0<u \le u_0/2$. For
this,  we let $Z^\mu$ be a future directed null vector which,
near the light-cone, takes the form $Z^\mu = a \partial_u +
r\partial_r$. Then
\bean
 T_{\mu\nu} X^\mu n^\nu & = &   ( uT_{u\nu} +  rT_{r\nu} )
 n^\nu
\\
 \nonumber
 & = &  \left(  ( u - a)T_{u\nu} +  aT_{u\nu} +  rT_{r\nu} \right)
 n^\nu
\\
 \nonumber
 & = &   (u-a) T_{0\nu}n^\nu
 +  \ub{ T_{\mu\nu} Z^\mu
 n^\nu}_{\ge 0}
 \;.
\eeal{5II9.5}
The last term is non-negative by the dominant energy condition.
\Eq{Tnx} will follow if
\bel{8IV.1}
 (u-a)\ge \frac 12 (t-r)
 \quad \Longleftrightarrow \quad a \le  \frac 12 (t-r)
 \;.
\ee
Now, the condition that $Z^\mu$ is  null  reads:
\bel{5II9.0}
 0= g_{\mu\nu}Z^\mu Z^\nu =  g_{uu} a^2 +
2g_{ur} a r  +r^2  g_{rr}
 \;. \ee
Keeping in mind that $g_{ur}$ is negative, we choose the
solution
$$
 a = \frac r {g_{uu}} \left(-g_{ur}-\sqrt{g_{ur}^2 -g_{uu} g_{rr}}\right)
= \frac {r g_{rr}}{-g_{ur}+\sqrt{g_{ur}^2 -g_{uu} g_{rr}}
 }
 \;.
$$
As already seen, since $g_{rr}=0$ at $u=0$, the hypothesis
$\partial_u g_{rr}|_{u=0}=0$ implies $|g_{rr}|\le Cu^2$ on any
bounded domain of $u$ and $r$, from which it easily follows
that, reducing $u_0$ if necessary, for $0\le u \le u_0$ the
inequality \eq{8IV.1} holds.

\medskip

We continue with:

\begin{Proposition}
On any bounded interval of $t$, say $0\le t \le T$,  and
assuming as before that we are working within the domain of
definition of normal coordinates, there exists
a constant $C$ such that, for $0<r\le t\le T$,
\bea
 \Big|\mcL_X g_{\mu\nu}-\Big(2 g_{\mu\nu}
   -\frac {\partial_r \alpha}{r}  X_\mu X_\nu
 \Big)\Big
 |_b \le C (t-r)
   \;,
\eeal{17III.6}
where the norm $|\cdot|_b$ is taken with some arbitrarily
chosen Riemannian metric $b$.
\end{Proposition}

\proof For $u\ge u_0/2$ the estimate is clear, so it remains to
consider the region $0\le u\le u_0/2$, where $X$ takes the form
\eq{10IV.1}. By definition of Lie derivative,
\bel{10II.25x}
 \mcL_X dz^\mu |_{u=0}  =
 \delta^\mu_r dr +  \delta^\mu_u du
 \;.
\ee
Writing the metric along the light-cone as $g=\eta +
(1-\alpha)du^2$, with $\eta=-du^2-2du dr + r^2 \mathring
h_{AB}dx^A dx^B$, one obtains
\bel{}
 \mcL_X \eta = 2 \eta , \qquad
 \mcL_X \left[(1-\alpha)du^2\right] = -r\partial_r\alpha\, du^2
  + 2(1-\alpha)du^2 ,
\ee
and, still at $\{u=0\}$,   one finds
\beal{17III.2}
 \mcL_X g &=& 2 g
   -r\partial_r\alpha\,du^2 
 \;.
\eea
We note the estimate
\bel{5II.15} 
|r\partial_r \alpha|\le Ct^2 \ \mbox{ for }\ 0< r \le t \le T
 \;.
\ee
On the light cone we have
$$
 du =  -\frac 1 r X_\mu dx^\mu
 \;.
$$
%
So we can rewrite \eq{17III.2} as
\beal{17III.4}
 \mcL_X g_{\mu\nu}|_{u=0} &=& 2 g_{\mu\nu}
   -\frac 1 r\partial_r\alpha\ X_\mu X_\nu  
 \;.
\eea
A Taylor expansion at $u=0$ gives \eq{17III.6}.
 \qed

\medskip

Let $E(s)$ be defined as in \eq{5II.10}, except that $t$ there
is taken now to be a normal coordinate $x^0$ within its domain
of definition. Recall that
\bel{5II.11}
 \Omega(s):= J^+(p)\cap \{t < s\}
  \;.
\ee
We consider the divergence  identity on $\Omega(s)$:
\bean
  E(s)
 +\int_{\mcC(s)} T^\mu{}_\nu X^\nu dS_\mu
 & = & - \int_{\partial \Omega(s)}T^\mu{}_\nu X^\nu dS_\mu=
  - \int_{  \Omega(s)}\nabla_{\mu}(T^\mu{}_\nu X^\nu )
\\
 & = & - \int_{  \Omega(s)} \frac{1}{2} T^{\mu\nu} \mcL_X g_{\mu\nu}
 \;.
\eeal{5II.12}
Since $T_{\mu\nu}$ is traceless by hypothesis, from
\eq{17III.6} and from  \eq{LDEC1} we obtain
\bel{5II9.13}
 |T^{\mu\nu} ( \mcL_X g_{\mu\nu}     -\frac{\partial_r \alpha}r X_\mu X_\nu
  )|_b
  \le C   (t-r) T_{\mu\nu}n^\mu n^ \nu
 \;.
\ee
As $\partial_r \alpha/r$ is bounded, \eq{5II9.13} 
together with \eq{LDEC2} imply
\bean
 |T^{\mu\nu}   \mcL_X g_{\mu\nu} |
  & \le &
    C \Big(   T_{\mu\nu}X^\mu X^ \nu
  +     (t-r) T_{\mu\nu}n^\mu n^ \nu\Big)
\\
  & \le &
    C' \Big(   T_{\mu\nu}n^\mu X^ \nu
  +   (t-r) T_{0\nu} n^ \nu\Big)
 \;.
\eeal{5II9.14}
By \eq{Tnx} the right-hand-side is bounded by a multiple of
$T_{\mu\nu}n^\mu X^ \nu$, and we can conclude that
$$
 E(s) \le C \int _0 ^s E(t) dt - \underbrace{\int_{\mcC(s)} T^\mu{}_\nu X^\nu dS_\mu}_{=0}
 \;,
$$
where the vanishing of the last integral follows from the fact
that $X^\nu$ is tangent to the generators of $\mcC$, hence null
there, and from \eq{3XII8.11}. Since $E(s)  $ approaches zero
as $s$ tends to zero, from Gronwall's lemma we obtain
$$
 \mbox{$E(s)=0$ for $0<s<s_*$.}
$$
Positivity of the integrand implies
\bel{4XII8.8x}
 T_{\mu \nu} X^\mu n^\nu =0 \ \mbox{ on
 $\Omega(s_*)$}
 \;.
\ee
Since $X$ is timelike on the interior of $\Omega(s_*)$, from
\eq{LDEC2} we conclude that the space-time is vacuum in
$\Omega(s_*)$.

\section{$g^{AB}R_{AB}$}
 \label{AppCheck2}
In this appendix we continue our analysis for a metric which,
in addition to  the hypotheses of Appendix~\ref{AppCheck},
satisfies further $\nu_A=0$ at $\{u=0\}$; thus, there we have
\bel{Axcharmetc}
 g = -\alpha du^2 + 2 \mysign  du dr +
 h_{AB} dx^A dx^B
 \;,
\ee
with
 \bel{Axinvcharmet}
 g^\sharp = \alpha \partial^2_r +
  2 \mysign \partial_u \partial_r + h^{AB}\partial_A
 \partial_B
  \;.
\ee

As in Appendix~\ref{AppCheck}, all calculations are done on the
null hypersurface $\{u=0\}$.

In addition to the previous list of vanishing Christoffel
symbols,
\bel{AxHGvanlist}
 \Gamma^u_{rr}=
 \Gamma^A_{rr}=
 \Gamma^u_{rA} =0
  \;,
  \ee
we now also have, due to the wave-map conditions and the
vanishing of $\nu_A$,
 \bea
 & \displaystyle
   \partial_u  g_{rr}=
   \partial_ u g_{r A}=
   \Gamma^u_{ur }=
   \Gamma^u_{uA}=
   \Gamma^r_{rr}=
   \Gamma^r_{r A}=
   \Gamma^A_{ru}=0
 \;. &
 \eeal{AxHGvanlist2}
The remaining Christoffel symbols can be obtained from those
listed in appendix \ref{AppCheck} by setting $\nu_A=0$ there.

We will need the following traces:
\bea
 \Gamma^\alpha_{u\alpha} &=&\frac{1}{2r^2} \zh^{AB}\partial_u g_{AB}
 +\mysign \partial_u g_{ru}
 \;,
\\
 \Gamma^\alpha_{r\alpha} &=& \frac{n-1}{r}
 \;,
\\
 \Gamma^\alpha_{A\alpha} &=&\frac {1}{2} \zh^{BC}\partial_A \zh_{BC}
 \;.
\label{Axgabc} \eea

In view of \eq{12I.2x},
\bel{10II.21xz}
 \lambda^r = \frac{\mysign}{2r^2}\zh^{AB}\partial_u g_{AB}
          + \frac{n-1}{r}\alpha + \partial_r \alpha
 \, , \qquad \zlambda^r = \frac{n-1}{r} - 2\Lambda r
 \frac{n+1}{n(n-1)}
  \; ,
\ee
and  of  the wave-map condition $\lambda^r = \zlambda^r$, we
conclude that
\be \label{ANRx3x} \frac{\mysign}{2r^2}\zh^{AB}\partial_u
g_{AB} = \frac{n-1}{r}(1-\alpha) - 2\Lambda
r\frac{n+1}{n(n-1)}-
\partial_r \alpha \, . \ee

We want, next, to calculate
\bean
   g^{AB} R _{A  B}
\nonumber
 & =&
 g^{A B} (\partial_\alpha
\Gamma^\alpha_{A B}-\partial_B \Gamma^\alpha_{A\alpha} +
    \Gamma^\alpha_{\sigma\alpha}\Gamma^\sigma_{A B}-
\Gamma^\alpha_{\sigma B}\Gamma^\sigma_{A \alpha})
 \\
 \nonumber
 & =&
 g^{A B} \Big(\partial_u
\Gamma^u_{A B}+ \partial_r \Gamma^r_{A B}
 + \partial_C \Gamma^C_{A B}
 - \partial_B \Gamma^\alpha_{A\alpha}
 +
    \Gamma^\alpha_{u\alpha}\Gamma^u_{A B}
 \\
 &   & +
    \Gamma^\alpha_{r\alpha}\Gamma^r_{A B}
    +
    \Gamma^\alpha_{C\alpha}\Gamma^C_{A B}
 -
    \Gamma^D_{C B}\Gamma^C_{A D}-
    2
    \Gamma^r_{C B}\Gamma^C_{A r}
 - 2
    \Gamma^u_{C B}\Gamma^C_{A u}\Big)
 \;.
 \phantom{xxxxxx}
\eeal{xANR1}
We will calculate separately various terms above, in random
order, starting with the last two:
\bea
 -2h^{A B}
    \Gamma^C_{u A}\Gamma^u_{B C}
 & = &
    \frac{\mysign}{r^3}   {
\zh^{AB}} { \partial_u g_{AB}}
 \;,
\\
 -2h^{A B}
    \Gamma^C_{r A}\Gamma^r_{B C}
 & = &
    \frac{\mysign}{r^3}   {
\zh^{AB}} { \partial_u g_{AB}} +  \frac{2(n-1)}{r^2} \alpha
 \;,
\\
 h^{A B} \Big(\partial_u
 \Gamma^u_{A B}+ \partial_r \Gamma^r_{A B} \Big)
 & = &
\nonumber
 -\frac{\mysign}{r^2}{\zh^{A B} {\partial_r\partial_u g_{AB}}}
\\
&&
 - \frac{n-1}{r^2} \partial_r(r \alpha)
 + \frac{n-1}{r} \partial_u g_{ur}
 \;,
\label{xANR3} \eea
\bean
 \lefteqn{  h^{A B} \Big( \Gamma^\alpha_{u\alpha}\Gamma^u_{A B} +
     \Gamma^\alpha_{r\alpha}\Gamma^r_{A B}\Big)}
 &&
\\
\nonumber
 &&=
    -\mysign \frac{n-1}{r^3} \zh^{AB}\partial_u g_{AB}
  - \frac{(n-1)^2}{r^2}\alpha
-\frac{n-1}{r}\partial_u g_{ur}
 \;.
\eeal{xANR4}
The remaining terms are
\beaa
 {
 g^{A B} \Big(  \partial_C \Gamma^C_{A B}
 - \partial_B \Gamma^\alpha_{A\alpha}
    +
    \Gamma^\alpha_{C\alpha}\Gamma^C_{A B}
 -
    \Gamma^D_{C B}\Gamma^C_{A D} \Big)
    }  = h^{AB}{\cal R}_{AB}
 \;,
\eeaa
where $\cal R$ is the Ricci tensor of the metric $h_{AB}$.
Adding, one is led to the simple identity
\bean
   g^{AB} R _{A  B}
 & =   &
-\frac{\mysign}{r^2}\zh^{AB}\partial_r\partial_u g_{AB}
-\frac{\mysign}{r^3}(n-3)\zh^{AB}\partial_u g_{AB}
\\
 &&
 - \frac{(n-1)(n-3)}{r^2}\alpha -
 \frac{n-1}{r^2}\partial_r(r\alpha)
   + h^{AB}{\cal R}_{AB}
 \;.
\eeal{xANR11}
But, in view of the wave-map condition \eq{ANRx3x},
\bean
 \lefteqn{-\frac{\mysign}{r^2}\zh^{AB}\partial_r\partial_u
 g_{AB} -\frac{\mysign}{r^3}(n-3)\zh^{AB}\partial_u g_{AB} }
 &&
\\
 && =
4\Lambda\frac{n+1}{n-1} +2\partial_r\partial_r\alpha
+\frac{4(n-1)}{r}\partial_r\alpha\
+\frac{2(n-1)(n-2)}{r^2}(\alpha-1) \; .
 \phantom{xxxxxx}
\eeal{ANRx3xy}
For the model metrics \eq{10II.10} we have $h^{AB}{\cal R}_{AB}
= (n-1)(n-2)/r^2$, so
\begin{equation}
g^{AB}R_{AB} =
   4\Lambda\frac{n+1}{n-1} +
   2 \partial_r\partial_r \alpha + \frac{3(n-1)}{r} \partial_r\alpha
   + \frac{(n-1)(n-2)}{r^2}(\alpha-1) .
\end{equation}

\bigskip

\ack {PTC and YCB are grateful to the Mittag-Leffler Institute,
Djursholm, Sweden, for hospitality and financial support during
the major part of work on this paper. They acknowledge useful
discussions with  Vincent Moncrief, as well as comments from
Roger Tagn\'e Wafo.  YCB  wishes to thank Thibault Damour for
making available his detailed manuscript calculations leading
to equations (22) of \cite{DamourSchmidt}. JMM thanks OxPDE for
hospitality. He was supported by the French ANR grant
BLAN07-1\_201699 entitled ``LISA Science'', and also in part by
the Spanish MICINN project FIS2008-03221. PTC was supported in
part by the EC project KRAGEOMP-MTKD-CT-2006-042360, and by the
EPSRC Science and Innovation award to the Oxford Centre for
Nonlinear PDE (EP/E035027/1).}

\section*{References}\def\cprime{$'$} \def\cprime{$'$} \def\cprime{$'$} \def\cprime{$'$}

\end{document}